\documentclass[twoside]{article}
\usepackage{flushend}

\usepackage{graphicx}
\usepackage{subfigure}
\usepackage{rotating}

\usepackage{amsmath}
\usepackage{amssymb}

\usepackage{indentfirst}
\usepackage{url}

\usepackage{subfigure}
\usepackage{multirow}
\usepackage{wrapfig}

\usepackage[ruled,vlined,linesnumbered,commentsnumbered]{algorithm2e}

\newcommand{\GET}{$\leftarrow$}

\newcommand{\intf}{\mathcal{I}}
\newcommand{\sched}{\mathcal{S}}

\usepackage[utf8x]{inputenc}

\usepackage{footmisc}

\usepackage[usenames]{color}

\newcommand{\ieee}{{\sc ieee} 802.11\xspace}

\begin{document}

\title{\vspace{-2cm}Broadcast Strategies with Probabilistic Delivery Guarantee in Multi-Channel Multi-Interface Wireless Mesh Networks}

\author{
Carina Teixeira de Oliveira 
\vspace{0.4cm}\\
{\small Grenoble Informatics Laboratory UMR 5217}
\\
{\small  UJF-Grenoble 1-CNRS}
\\
{\small 681 Rue de la Passerelle, BP72}
\\
{\small 38402 Saint Martin d'Heres, France}
\\
{\small Email: \url{oliveira@imag.fr}}
\and Fabrice Theoleyre
\vspace{0.4cm}\\
{\small CNRS -- LSIIT  -- UMR 7005}
\\
{\small University of Strasbourg}
\\
{\small Boulevard Sebastien Brant}
\\
{\small 67412 Illkirch, France}
\\
{\small Email: \url{theoleyre@unistra.fr}}
\\[0.5cm]
\and Andrzej Duda
\vspace{0.4cm}\\
{\small Grenoble Informatics Laboratory UMR 5217}
\\
{\small  Grenoble Institute of Technology-CNRS}
\\
{\small 681 Rue de la Passerelle, BP72}
\\
{\small 38402 Saint Martin d'Heres, France}
\\
{\small Email: \url{duda@imag.fr}}
}

\date{}

\maketitle

\begin{abstract}
Multi-channel multi-interface Wireless Mesh Networks permit to spread the load across orthogonal channels to improve network capacity. Although broadcast is vital for many layer-3 protocols, proposals for taking advantage of multiple channels mostly focus on unicast transmissions. In this paper, we propose broadcast algorithms that fit any channel and interface assignment strategy. They guarantee that a broadcast packet is delivered with a minimum probability to all neighbors. Our simulations show that the proposed algorithms efficiently limit the overhead.
\end{abstract}



\section{Introduction}\label{sec:Introduction}

Wireless Mesh Networks (WMN) have attracted increasing attention in recent years
because of their
low-cost and ease of deployment.  
These multi-hop networks are self-organized and self-configured without any centralized control.
They are composed of static wireless routers and some of them act as gateways toward the Internet. 
In this paper, we consider WMN with routers based on the \ieee technology.

When mesh routers use a single interface (i.e. wireless network card) tuned to a single channel,
the network capacity degrades with the increase of the network size due to channel contention
and spatial problems such as exposed and hidden nodes~\cite{chaudet05}.
One way of improving the performance of WMN is to use multiple non-overlapping channels (free of inter-channel interference) so that mesh routers can transmit in parallel and without collisions~\cite{li09}.  
To take advantage of multiple channels, nodes may have multiple interfaces to simultaneously transmit/receive packets. 
These networks are called Multi-Channel Multi-Interface (MCMI) WMN.

It has been shown that one can significantly improve the network capacity by carefully choosing
the set of channels a mesh router may use for each of its interfaces and which of them it will use to communicate with its neighbors.
A pair of nodes should have sufficient channels in common with each other while minimizing interference with other active pairs.

Recently, there has been an increasing interest in developing new solutions
for MCMI WMN. 
However, few studies considered efficient broadcast in such networks---the work done has only focused on static interface assignment in which all nodes are tuned to 
a common set of channels \cite{qadir07,yang09}. 
While it is true that static approaches provide suitable stability
for routing protocols without path changes, re-ordering,
channel switches etc., on the other hand, these approaches
do not efficiently distribute the load among all available
channels. Consequently, the inability to adapt interfaces under heavy load
and interference variations can drastically reduce the overall
network performance.
Therefore, the channel and interface assignments play a key role on MCMI WMN performance, 
which can not be neglected by broadcast solutions.

We propose to focus here on the broadcast problem in multi-channel multi-interface wireless mesh networks.
Our contribution is threefold:
\begin{enumerate}
	\item we introduce a classification of Channel and Interface Assignment (CIA) strategies in MCMI WMN;

	\item we propose broadcast algorithms that fit any CIA strategy guaranteeing a packet is delivered with a minimum probability to all neighbors; 
	
	\item we provide simulation results to compare different strategies and choose
the most suitable one for a given situation. 
\end{enumerate}

The rest of this paper is organized as follows.
We first introduce the system model in the next section.
Section~\ref{section:related_work} presents a review of relevant related work.
In Section~\ref{section:probGuarantee}, we introduce the probabilistic delivery guarantee.
Section~\ref{section:algorithms} presents the broadcast algorithms with the
probabilistic guarantee. 
Simulation results are presented in Section~\ref{section:performance_evaluation}.
Finally, we conclude the paper in Section~\ref{section:conclusion}.


\section{Model and Assumptions}
\label{sec:model}

We model a wireless mesh network as an undirected graph $G=(V,E)$, where $V$ is the set of nodes (mesh routers) and $E$ the set of edges corresponding to two nodes able to directly communicate. 

We adopt the following notation:
\begin{itemize}
	\item $C$: the number of non-overlapping (orthogonal) channels that can be used by all nodes $v \in V$. 
	We assume that a node can tune each of its interfaces to a set of channels $c \subseteq {C}$;

	\item $Intf(v)$: the set of interfaces of node $v \in V$;

	\item $\sched(i)$: the schedule of interface $i$---a list of tuples $\{(channels,timeStart,timeStop)\}$.
	The schedule may be periodical or anarchical;

	\item $\intf_{v}$: the number of interfaces of node $v$, where $\intf_{v}= \intf_{v}^S+\intf_{v}^D$ with $\intf_{v}^S$ denoting the number of static interfaces and $\intf_{v}^D$ the number of dynamic interfaces of node $v  \in V$.

\end{itemize}

\begin{figure}[t!]
\centering
	\subfigure[Static.] { 
		\includegraphics[width=0.45\textwidth]{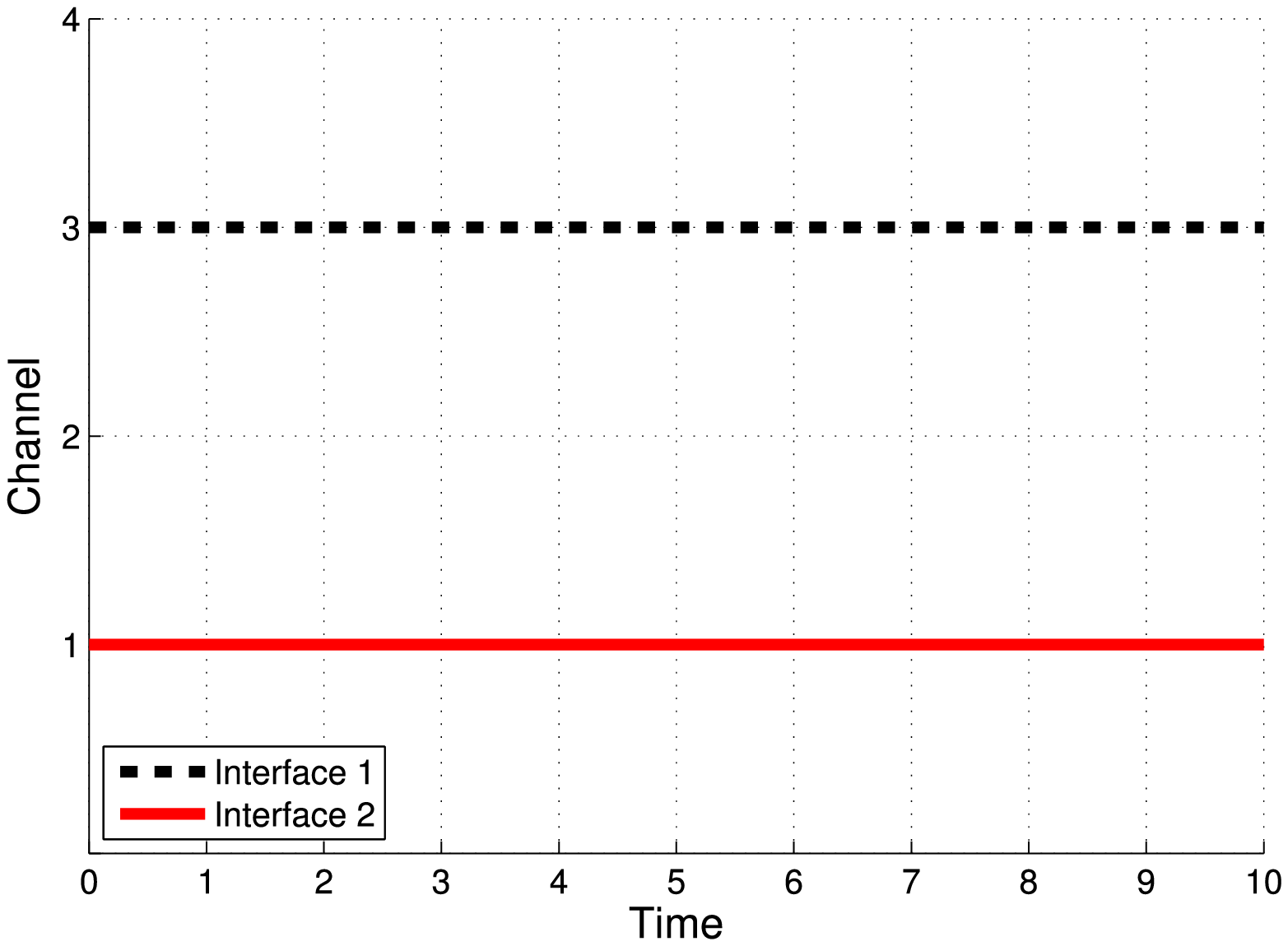}
		\label{fig:StaticInterfaceAssignment}
	}
	\subfigure[Dynamic.] { 
		\includegraphics[width=0.45\textwidth]{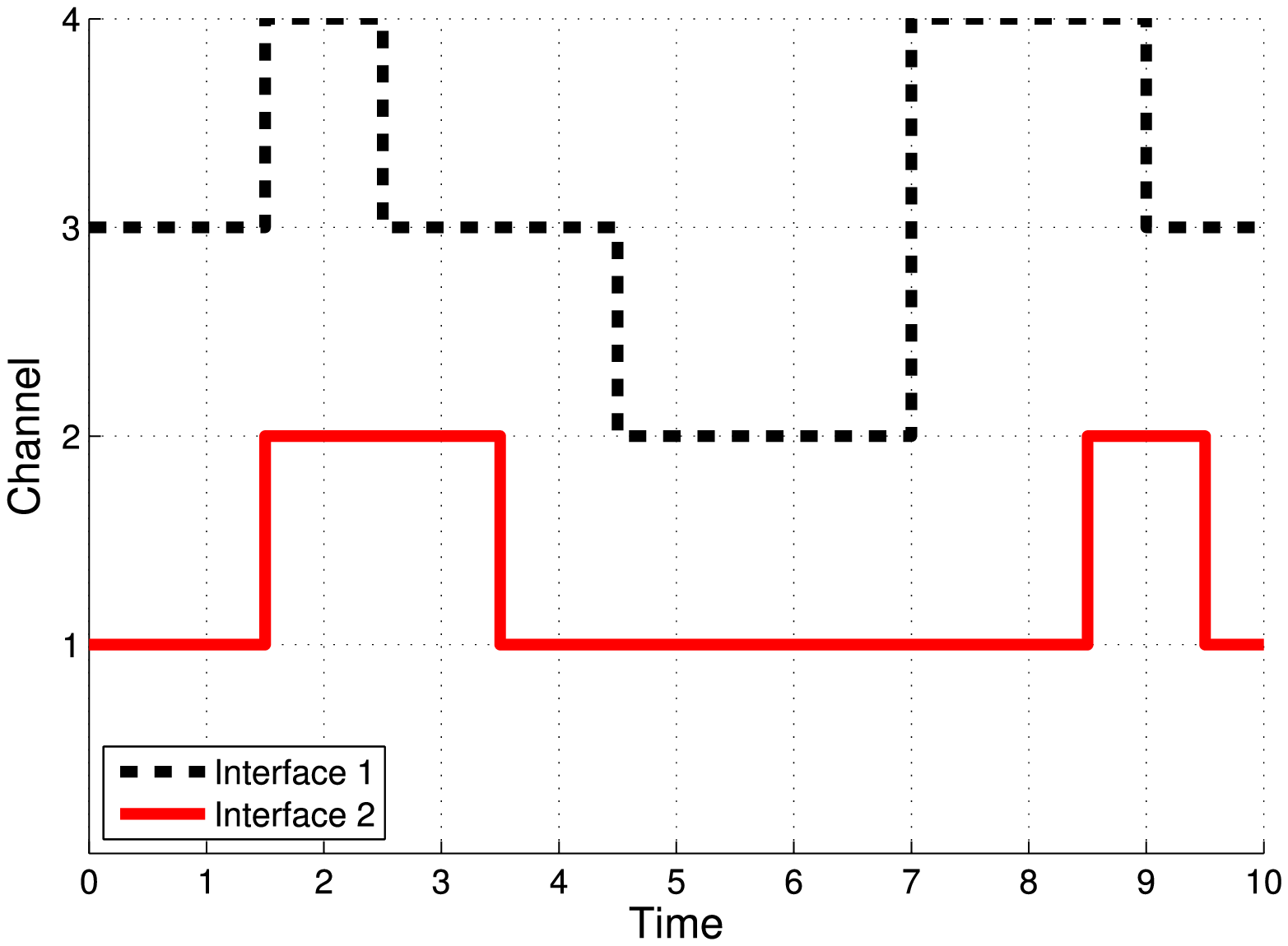}
		\label{fig:DynamicInterfaceAssignment}
	}
	\subfigure[Mixed.] { 
		\includegraphics[width=0.45\textwidth]{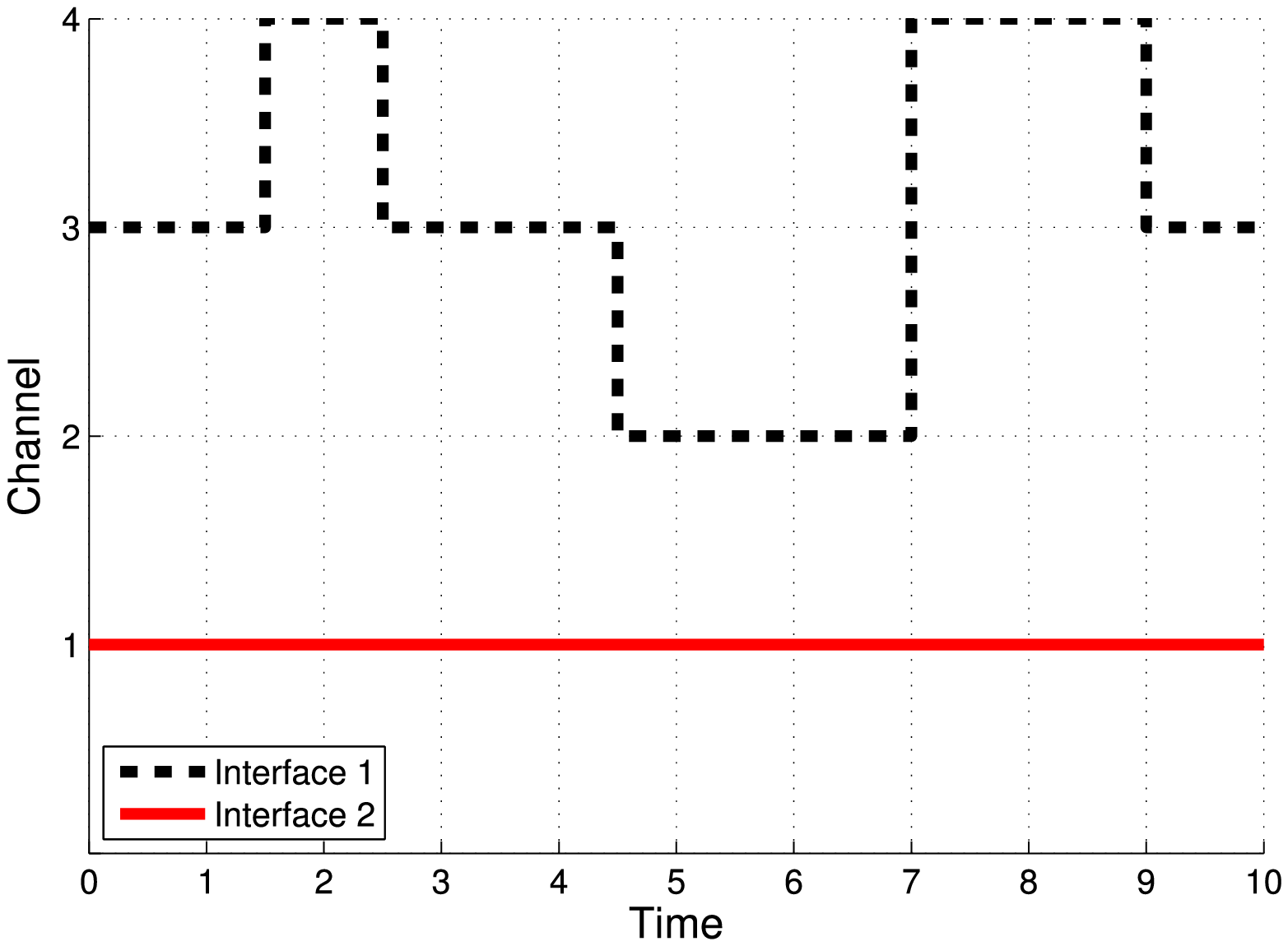}
		\label{fig:MixedInterfaceAssignment}
	}
 	\caption{Examples of interface assignment with 4 channels and 2 interfaces.}
	\label{fig:interface-assignment}
\end{figure}

Two mesh routers in the transmission range of each other are able to communicate if at least one of their interfaces uses the same channel at the same instant.
More formally, $(u,v) \in V^2, (u,v) \in E$ if:
\begin{equation}
	\exists i \in Intf(u), \exists j \in Intf(v) \ such\ that\ 
	\sched(i) \cap \sched(j) \neq \emptyset
\end{equation}


\subsection{Interface Assignment}
\label{section:InterfaceAssignment}

We distinguish between three types of \textit{interface assignment}:

\subsubsection{Static Interfaces}

In this case, all interfaces are static and remain on a channel for a long period of time:
\begin{equation}
 \forall v\in V, \; \intf_{v}^D=0\rightarrow \intf_{v}= \intf_{v}^S.
\end{equation}

Figure \ref{fig:StaticInterfaceAssignment} shows an example of static interface assignment with four channels and two interfaces.
The $x$-axis is the time and the $y$-axis is the channel id.
Note that each interface remains on the same channel regardless of time.

\subsubsection{Dynamic Interfaces} 
In this case, all interfaces are dynamic and frequently switch from one channel to another:
\begin{equation}
 \forall v\in V, \; \intf_{v}^S=0\rightarrow \intf_{v}= \intf_{v}^D.
\end{equation}

Figure \ref{fig:DynamicInterfaceAssignment} presents an example of dynamic interface assignment.
Note that two interfaces of a node should not use the same channel at the same instant.
Otherwise, we would waste bandwidth since the pair of interfaces will interfere.

\subsubsection{Mixed Interfaces} 
\label{sec:mixed} 
In this case, $\intf_{v}^S$ static interfaces permanently stay on a channel and $\intf_{v}^D$ dynamic interfaces frequently switch from one channel to another:
\begin{equation}
 \forall v\in V, \; \intf_{v}^S\geq 1, \; \intf_{v}^D\geq 1.
\end{equation}

Figure \ref{fig:MixedInterfaceAssignment} shows an example of mixed interface assignment in which \textit{Interface 1} switches from one channel to another and \textit{Interface 2} remains tuned to channel 1.


\subsection{Channel Assignment}
\label{section:ChannelAssignment}

Furthermore, we classify \textit{channel assignment} according to one of the following approaches:

\subsubsection{Common Channel Set} 

Static interfaces correspond to a \textit{Common Channel Set} (CCS)---once a channel is assigned to an interface, it does not change. 

	One of the simplest set works as follows: on every node,  interface 1 is assigned to channel 1, interface 2 is assigned to channel 2, and so on. 
	CCS can also be built upon the concept of non-overlapping channels in order to reduce interference. 
	Adya \textit{et al}. simulated scenarios in which each node has two static interfaces tuned to orthogonal channels: one interface is assigned to channel 1 and the other one to channel 11~\cite{adya04}.

\subsubsection{Pseudo-Random} 

Static interfaces at different nodes may be assigned to a different set of channels. Each node locally decides which channel to allocate to static interfaces. 

  A simple solution consists in choosing channels at
  random~\cite{mao07}. Another solution is to use some well-known
  function $f$ of the node identifier to select the channels to assign to the
  static interfaces~\cite{kyasanur05}. Neighbors of a node $v\in V$ can use the
  same function $f$ to compute the channel used by $v$. An alternative solution
  consists in explicitly exchanging \textit{hello} packets that contain
  information about the channels used by static interfaces. Based on the
  received \textit{hello} packets, nodes may choose to set their channels to an
  unused or a lightly loaded channel.

\subsubsection{Adaptive}

It uses some information or criteria to dynamically
  assign channels based on for instance time instants (defined or random), a
  pre-defined channel visiting order, the interference level, or the available
  bandwidth.

Finally, we distinguish between the following types of broadcast communications:
\begin{itemize}
 	\item Discovery Broadcast: a node has to discover neighbors;
 	\item Local Broadcast: all neighbors must receive a packet;
 	\item Flooding: a packet has to be received by all nodes in the network.
\end{itemize}

We will focus here on  Local Broadcast. 
We considered  Discovery Broadcast elsewhere \cite{abdelali10}.


\section{Related Work}
\label{section:related_work}

\subsection{Multi-channel Multi-interface Strategies}
\label{section:strategies}

\begin{table*}[ht]
 \caption{Channel and Interface Assignment (CIA) Strategies}
 \centering {
  \begin{tabular}{|l|c|c|c|c|c|c|c|c|}
   \cline{1-8} \multicolumn{1}{|c|}{\multirow{2}{*}{\textbf{Strategies}}}&\multicolumn{3}{c|}{\textbf{Interface}} &\multicolumn{3}{c|}{\textbf{Channel}} &\multicolumn{1}{c|}{\multirow{2}{*}{\textbf{Ref}}}\\
   \cline{2-7} &S &
   D &
   M &%
   C &
   {P-R} &
   A &\\
   \hline  \textit{Static/Common} &X&%
   &%
   &%
   X&%
   &
   &
   \cite{adya04,draves04} \\
   \hline \textit{Static/Pseudo-Random}&X &%
   &%
   &%
   &%
   X&
   &
   \cite{mao07,marina09} \\ 
   \hline  \textit{Dynamic/Adaptive}&&%
   X&%
   &%
   &%
   &%
   X	&
   \cite{dasilva08,bahl04} \\
   \hline  \textit{Mixed/Common \& Adaptive} & &%
   &%
   X&%
   X&%
   &%
   X	&
   \cite{so04,tseng01} \\
   \hline \textit{Mixed/Pseudo-Random \& Adaptive} &&%
   &%
   X&%
   &%
   X &%
   X 	&
   \cite{kyasanur05,ding09}\\
   \hline
  \end{tabular}
 }
Interface Assignment -- S: Static, D: Dynamic, M: Mixed \\
Channel Assignment -- C: Common, P-R: Pseudo-Random, A: Adaptive
\label{tab:strategies}
\end{table*}

A multi-channel multi-interface strategy consists of a combination of
\textit{interface} and \textit{channel} assignments \cite{si10}. 
We can identify the following strategies:

\begin{enumerate}
	\item \textit{Static Interfaces/Common Channel Assignment:}
all interfaces are static and use the CCS approach (all nodes use channel $i$ on
the $i^{th}$ interface)~\cite{adya04,draves04}. 

We can note that the strategy is optimal when the number of interfaces equals the number 
of available channels, which is seldom the case;

\item \textit{Static Interfaces/Pseudo-Random Channel Assignment:}
similar to the previous strategy, it assigns a channel to each interface for permanent use. 
However, static interfaces at different nodes may be assigned to a different set of channels~\cite{mao07,marina09}.

This approach does not guarantee connectivity since two nodes may choose different channels for their interfaces. Therefore, deafness may arise;

\item \textit{Dynamic Interfaces/Adaptive Channel Assignment:}
all interfaces are dynamic. Thus, the topology is also dynamic:
two nodes may temporarily be able to communicate until one interface switches its channel. 
Often, a rendezvous mechanism has to be used to avoid deafness~\cite{dasilva08,bahl04}. 
In other words, nodes can agree on the channels they will use in the next time interval. Then, channel 
re-assignment ensures that statistically, a node can reserve a common channel for each of its neighbors
after a sufficiently long time.

\item \textit{Mixed Interfaces/Common and Adaptive Channel Assignment:} 
static interfaces use a common channel set (CCS) while dynamic interfaces act in an on-demand manner~\cite{so04,tseng01}.

A pair of nodes may use its static interfaces to negotiate a channel for data
exchange. 
Negotiation takes place on a dedicated \textit{control channel} to isolate
control RTS/CTS traffic from data.

The strategy uses all available channels with a trade-off between connectivity (static interfaces) 
and capacity optimization (dynamic interfaces);

\item \textit{Mixed Interfaces/Pseudo-Random and Adaptive Channel Assignment:}
\label{sec:strategy5-definition}
different nodes assign their static interfaces to different channels, while the remaining interfaces switch channels in an adaptive manner.
Static interfaces are used in reception while dynamic interfaces are used for transmissions. 
A node has just to know the list of channels used by the static interfaces of its neighbors to communicate with them: no deafness appears. 

This strategy often uses one single static interface ($\intf_{v}^S=1$) \cite{kyasanur05}.

\end{enumerate}

Table~\ref{tab:strategies} presents an overview of all Channel and Interface Assignment (CIA) strategies.

\subsection{Broadcast}

Many network protocols use local broadcast for various purposes: routing, coordination, synchronization, etc.
The functionality of reaching all neighbors in a local broadcast is useful in flooding and efficient flooding protocols aim at limiting the number of retransmissions.  
In a single channel network, one single packet is sufficient to \emph{cover} all the neighbors (all neighbors receive it), because of the broadcast nature of radio transmissions.

Several papers tried to tackle the broadcast problem in multi-channel multi-interface wireless mesh networks.
Qadir \textit{et al.} \cite{qadir07} proposed to optimize the delay for multi-rate mesh networks. 
However, they focus on the global flooding problem, i.e. how each node in the network receives the flooded packets.
Each node selects a subset of neighbors to forward the packet, pruning redundant transmissions.
They assume a static interface assignment (Static/Common) in which all nodes are tuned to the same channels.
Song \textit{et al.} \cite{song07} presented also independently a broadcast protocol to achieve a 100\% reliability with minimum latency. 
They constructed a broadcasting tree using a link quality metric, but focused also on a network wide flooding. 
Yang \textit{et al.} \cite{yang09} additionally introduced network coding to reduce the associated overhead in mesh networks.

Xing \textit{et al.}  \cite{xing07b} proposed superimposed codes to tackle both the unicast and broadcast problems in multichannel multiradio mesh networks.

In conclusion, no proposal is sufficiently generic to deal with any CIA strategy.


\section{Probabilistic delivery guarantee} 
\label{section:probGuarantee}

Transmission in wireless networks may suffer from errors due to various effects
at PHY and MAC layers: attenuation, interference, fading, multipath propagation,
synchronization errors, or collisions. Our goal is to design broadcast protocols
that guarantee the reception of a broadcast packet by each neighbor with a
certain probability.

\subsection{Packet Error Estimation}

We denote by $p_e$ bit error probability and by $p_p$ packet error probability.
They are related by the following relation:
\begin{equation}
p_{p} = 1 - (1-p_e)^{size}
\end{equation}
where $size$ denotes the size in bits of a packet. $p_{deliv}$ is the
probability of successful packet delivery $p_{deliv} = 1 - p_p $.
Since it depends on a given radio link, we use the notation $p_{deliv}(u,v)$ for
the transmission from $u$ to $v$. 

\subsection{Probabilistic guarantee}

We propose a \textit{probabilistic guarantee} of local broadcasts. 

We consider that a particular neighbor is \emph{covered} by a broadcast if it
receives at least one copy of the corresponding packet with a probability
superior or equal to $p_{cover_{min}}$, a
parameter of the protocol.
Higher layers may specify its value when they want to transmit a broadcast
packet.
A local broadcast is \emph{successful} if all neighbors are \emph{covered}. 

Let $N(v)$ represent the neighbors of $v$.
We denote by $p_{cover}(u \rightarrow v)$ the probability that node $v$ correctly
receives the broadcast of node $u$, i.e., $v$ is \emph{covered}.
Our protocol will imply that
\begin{equation}
	\forall  v \in N(u), p_{cover}(u \rightarrow v) \geq p_{cover_{min}}.
	\label{eq:broadcast_successful}
\end{equation}

To provide guarantees, we limit the links to those with packet error
probability of at least $p_{p_{max}}$, i.e. a node does not maintain radio
links of low quality. 


\section{Broadcast Algorithms}
\label{section:algorithms}

In this section, we introduce the broadcast algorithms based on the classification
proposed in Section~\ref{section:strategies}.

\subsection{Static Interfaces with Common Channel Assignment}
\label{section:algo_ccs}

With the common channel assignment, the $i^{th}$ channel is assigned to the
$i^{th}$ static interface, so there is no deafness. 
Thus, broadcast is simple: a node has just to broadcast a packet through 
any of its static interfaces and all its neighbors will receive it.

A node has to send as many copies of the packet as required to cover each of its neighbors with the expected probability.
If we consider packet losses uncorrelated among the different copies, the probability the node $v$ receives at least one of the $k$ copies from $u$ is:
\begin{equation}
	p_{cover}(u \rightarrow v) = 1 -  \left(1 - p_{deliv}(u,v)\right)^k
	\label{eq:pcover_kcopies}
\end{equation}

Finally, a node has to send the following number of copies so that $v$ receives
the packet with a probability superior to
$p_{cover_{min}}$:
\begin{equation}
	K = \left\lceil \frac{log (1 - p_{cover_{min}})}{log(1 - p_{deliv}(u, v) ) } \right\rceil
	\label{eq:per_common_channel}
\end{equation}
The link with the smallest $p_{deliv}$ will determine the lower bound of the
number of copies to transmit. 

When only one static interface is tuned to the control channel, we can use this interface to send broadcast packets. 
However, the whole control traffic is concentrated on the control channel thus leading to its high utilization for
large broadcast load.

We can apply this approach to Strategies~1 and 4 (those that use a Common Channel Set) in Section~\ref{section:strategies}.

\subsection{Static Interfaces with Pseudo-Random Channel Assignment}

\begin{algorithm}[t]
\caption{Greedy Selection for Static Interfaces with Pseudo-Random Channel Assignment}
\label{algo:greedyStatic}
\SetKwFunction{getListNeighStaticIntf}{getListNeighStaticIntf}
\SetKwFunction{getNeighbors}{getNeighbors}
\SetKwFunction{rand}{rand}
\SetKwFunction{getNbNeighsCovered}{nbNeighsCovered}
\SetKwFunction{neighCovered}{neighCovered}
\SetKwFunction{sendBroadcast}{sendBroadcast}
\SetKwFunction{neighChannel}{neighUsing}
\SetKwData{neighs}{neighs}
\SetKwData{channels}{channels}
\SetKwData{bestChannel}{bestChannel}
\SetKwData{Pcover}{Pcover}
\SetKwData{Pdeliv}{Pdeliv}
\SetKwData{nbCovered}{nbCovered}
\tcc{initialization}
\tcc{neighs  $\equiv$ \{($neigh,channel$)\}}
\neighs[] \GET \getListNeighStaticIntf{} \;

\vspace{0.2cm}
\tcc{initially, no neighbor is covered}
\For{$i \in [0..| \neighs|]$}{
	\Pcover[i] \GET 0 \;
}
\vspace{0.2cm}
\tcc{if at least one uncovered neighbor exists}
\While{($\exists n \in \neighs $ such that $ \Pcover[n] < p_{cover_{min}}$)}{
	
	\vspace{0.1cm}
	\tcc{count the nb of neighbors covered for each channel.}
	\For{$c \in [0 .. |\channels |]$}{
		\nbCovered[c] \GET \getNbNeighsCovered{c} \;
	}	
	
	\tcc{select one of the best channels}
	\bestChannel \GET \rand{\nbCovered } \;
	
	\tcc{update the Pcover proba for each newly covered neighbor}
	\For{n $\in$ \neighChannel{\bestChannel} }{
		\tcc{this neighbor was not yet covered at all }
		\If{\Pcover[n] = 0}{
			\Pcover[n] $\leftarrow$ \Pdeliv[n] \;
		}
		\Else{
			\Pcover[n] $\leftarrow$ $1 - (1-\Pcover[n]) \cdot (1-\Pdeliv[n])$ \;
		}
	}
	\tcc{send one broadcast packet}
	\sendBroadcast{\bestChannel} \;
}
\normalfont
\end{algorithm}

With this kind of assignment, a single transmission is not sufficient for local broadcast, because not all neighbors use the same channel.
A node may have to send several packets so that all its neighbors become covered through different channels. 
In this strategy, each node knows the list of its neighbors and their static channels (this is a feature of the unicast protocol): a node will also use this information for its broadcast transmissions.

We propose a greedy approach inspired by MultiPoint-Relays \cite{MPR}: a node
chooses the minimum number of channels that cover the largest number of
neighbors (cf. Algorithm~\ref{algo:greedyStatic}).  
More precisely, a node proceeds in the following way:
\begin{itemize}
	\item a node constructs the list of its neighbors (i.e. all the nodes with which it has a common channel). 
	It initially considers that all its neighbors are \emph{uncovered} (Algorithm~\ref{algo:greedyStatic}, lines $1-6$).  
	
	\item while at least one neighbor is covered with a probability inferior
          to $p_{cover_{min}}$, the node searches for the channel with the best quality:
	\begin{itemize}
		\item it counts the number of newly covered neighbors for each channel (their covering probability is inferior to $p_{cover_{min}}$);
		\item it randomly chooses one of the best channels (to balance the load among channels);
		
		\item for each neighbor reachable through this channel, it updates the probability of reception.
		It corresponds to the delivery probability for the link $(u,v)$ if $u$ did not yet schedule a packet for $v$. Otherwise, it recursively applies Equation~\ref{eq:recursive_pcover}. 
	\end{itemize}	
\end{itemize}
\begin{eqnarray}
	p_{cover} (u,v) = 1 - \left(1 - p_{cover} (u,v)\right) \left(1- p_{deliv}(u,v) \right)
	\label{eq:recursive_pcover}
\end{eqnarray}

We can apply this approach to Strategies 2 and 5 in Section~\ref{section:strategies} that use static interfaces to receive packets.

\subsection{Dynamic Interfaces with Adaptive Channel Assignment}

\begin{algorithm}[t]
\caption{Greedy Selection for Dynamic Interfaces with Adaptive Channel Assignment}
\label{algo:greedyDynamic}
\SetKwFunction{constructSchedule}{constructSchedule}
\SetKwFunction{sort}{sort}
\SetKwFunction{sendBroadcast}{sendBroadcast}
\SetKwFunction{randMAX}{randAmongMAX}
\SetKwFunction{neighSlot}{neighDuring}
\SetKwData{Pcover}{Pcover}
\SetKwData{Pdeliv}{Pdeliv}
\SetKwData{schedule}{schedule}
\SetKwData{neighs}{neighs}
\SetKwData{slotId}{slotId}
\SetKwData{slots}{slots}
\SetKwData{bestSlot}{bestSlot}
\tcc{schedule $\equiv$ \{($T_{start},T_{end},neigh,channel$)\}: the channel used by each neighbor during each timeslot}
\schedule[] \GET \constructSchedule{\neighs}\;
\tcc{greedy steps}
\While{($\exists n \in \neighs $ such that $ \Pcover[n] < p_{cover_{min}}$)}{
	\tcc{sort the slots according to the number of new neighbors they cover}
	\slots \GET \sort(\schedule, \Pcover) \;
			
	\tcc{select one of the slots that covers most neighbors}
	\bestSlot \GET \randMAX{\slots } \;

	\tcc{update the Pcover probability}
	\For{n $\in$ \neighSlot{\bestSlot} }{
		\Pcover[n] $\leftarrow$ $1 - (1-\Pcover[n]) \cdot (1-\Pdeliv[n])$ \;	
	}
	
	\tcc{send one broadcast packet}
	\sendBroadcast{\bestSlot} \;

}
\normalfont
\end{algorithm}

When a node only uses dynamic interfaces, it needs to avoid deafness by correctly choosing both an interface and a schedule. 
We propose the Algorithm~\ref{algo:greedyDynamic}. 

\begin{figure}[t!]
\begin{center}
	\includegraphics[width=0.9\linewidth]{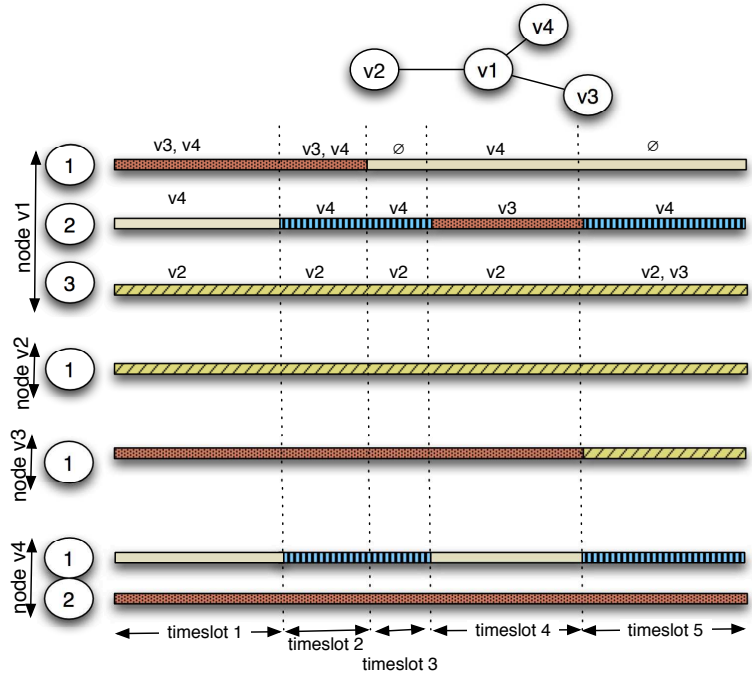}
	\caption{Local broadcast with mixed interfaces---\emph{each color of the bars represents a different channel and we report the list of neighbors reachable through each of $v_1$ interfaces at any instant ($v_1$ has 3 interfaces, $v_2$ and $v_3$ one interface, and $v_4$ 2 interfaces). We consider in this example a neighbor covered if it received at least one copy}}
 	\label{fig:broadcast}
	\end{center}
\end{figure}

A node first creates the schedule of its interfaces and thus of its neighbors:
a node knows the channel switching instants of all its neighbors for all their interfaces (otherwise transmissions are impossible due to deafness).
It constructs timeslots so that itself and all its neighbors stay tuned to the same channel during one timeslot.
We do not require all the nodes to switch their channels at the same time.
Let us consider the example in Figure~\ref{fig:broadcast} in which timeslots are delimited by dashed lines: the first interface of node $v_1$ stays tuned to the same channel during timeslots 1 and 2 while the second interface switches between both timeslots. 
The schedule consists of a kind of the lowest common denominator between the different channel switching instants for all neighbors.

After having constructed this schedule, the transmitter is able to compute the number of neighbors that can be covered for each interface for each timeslot (i.e. when the channels match).
Thus, it will re-iterate by greedily choosing pairs $<$\texttt{timeslot,interface}$>$ that cover the largest number of not yet covered neighbors. 
When a node sends a copy of a broadcast packet, it updates the probability of delivery for each neighbor, adopting the same approach as Algorithm~\ref{algo:greedyStatic} (lines $15-20$).
The algorithm stops when all the neighbors are covered with a probability superior to $p_{cover_{min}}$.

Let us consider the example in Figure~\ref{fig:broadcast}.
We consider in this example a neighbor covered if it received at least one copy.
As explained previously, $v_1$ first computes \emph{timeslots} (dashed lines).
Then, it chooses the neighbors reachable through each interface for each timeslot and applies the greedy algorithm. 
For instance, node $v_1$ can reach node $v_3$ during the first timeslot through its first interface and node $v_4$ through the second interface of $v_4$.
Finally, node $v_1$ may choose timeslot 1 via its first interface to cover $v_3, v_4$ and timeslot 1 via its third interface to reach node $v_2$.

This algorithm can apply to the strategy that only uses dynamic interfaces (Strategy 3 in Section~\ref{section:strategies}).


\section{Performance Evaluation} 
\label{section:performance_evaluation}

\begin{table}
	\caption{Default parameter values}
	\centering {
		\begin{tabular}{|p{5cm}|c|}
		\hline
		\textbf{Parameter}   			                  &  \textbf{Default value}\\
		\hline
		Number of nodes					& 200\\
		Density {\scriptsize (avg. number of neighbors)}	& 10\\
		\hline
		Number of interfaces					& 3\\
		Number of channels					& 12\\
		\hline
		$p_{cover_{min}}$	 {\scriptsize(probability above which we consider the node is covered)}	& 0.95\\
		\hline
		\end{tabular}
	}
	\label{table:parameters}
\end{table}

We have implemented a simulator to evaluate the broadcast performance\footnote{the simulator is freely available at \url{http://forge.imag.fr/projects/graphsim} in the subversion repository} \cite{simulator}. 
We assume the ideal MAC layer to focus on broadcast performance with no collision. 
We generate random positions of nodes and plot 95\% confidence intervals. 
We use the standard values depicted in Table~\ref{table:parameters}.
The simulation measures:
\begin{enumerate}
	\item the overhead defined as the average number of transmissions required by a node to cover all its neighbors;
	\item the Jain Index of the load for all the channels to measure fairness. Let $B_c$ denotes the bandwidth consumed by the broadcast on channel $c$:
	\begin{equation}
		JainIndex = \frac{\left(\sum_{c= 1}^{C} B_c\right)^2}{C \cdot \sum_{c = 1}^{C} B_c^2}
	\end{equation}
\end{enumerate}

We denote each strategy as introduced in Section~\ref{section:strategies} and apply the broadcast algorithms defined in the previous section.
In particular, we have implemented the Dynamic/Adaptive strategy in a way
that each interface equally shares its time among all the channels following a pseudo-random sequence \cite{bahl04}. 
Two nodes are able to exchange packets if at least one pair of interfaces uses the same channel at the same instant.

\subsection{Packet Error Rate}

\begin{figure}
\centering
	\includegraphics[width=0.7\linewidth]{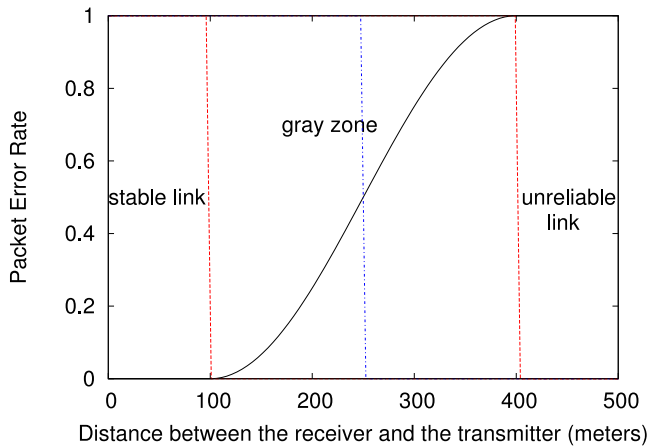}
 	\caption{Impact of the distance between the receiver and the transmitter on the Packet Error Rate (PER).}
	\label{fig:per}
\end{figure}

Simulation takes into account packet error probability through the Packet Error
Rate (PER) model represented in Figure~\ref{fig:per}  
\cite{camp06} that shows the relation between the PER and the distance.
As explained above, the neighbors with a PER superior to $p_{p_{max}}$ have not
to be covered. 
For the numerical results, we have chosen the value of $p_{p_{max}}=0.5$ although different values would lead to consistently the same results.

\subsection{Network cardinality}

\begin{figure}[t!]
\begin{center}
	\subfigure[Overhead]{
		\includegraphics[width=0.7\linewidth]{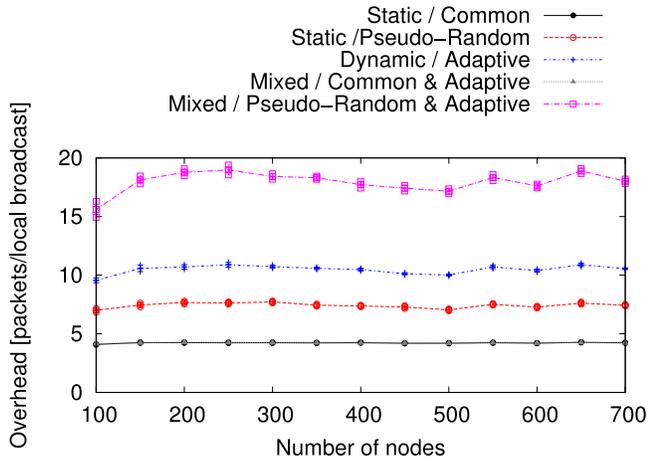}
		\label{fig:nbnodes_overhead}
	}
	\subfigure[Jain Index]{
		\includegraphics[width=0.7\linewidth]{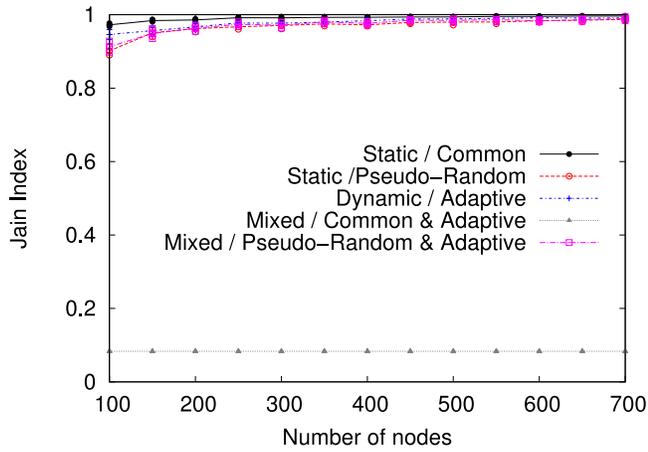}
		\label{fig:nbnodes_jain}
	}
	\caption{Impact of the number of nodes, 10 avg. neighbors, 3 interfaces, 12 channels}
\end{center}
\end{figure}

Figure~\ref{fig:nbnodes_overhead} presents the overhead in function of the number
of nodes when maintaining constant density. 
The Static/Common and Mixed/Common \& Adaptive strategies have the same minimal
overhead: no deafness arises and all neighbors receive a broadcast transmission. 
Since some neighbors may present a non-null Packet Error Rate, several non acknowledged broadcasts are required before considering they are \emph{covered}.

The Dynamic/Adaptive strategy shows a limited overhead by greedily choosing the most suitable channels.
The Static/Pseudo-Random strategy requires a little less broadcast packets (7 transmissions). 
Indeed, this strategy results in lower connectivity: two nodes may be in the radio range with each other, but may not share a common static channel. 
In this case, this kind of a \emph{virtual neighbor} is not anymore a
\emph{neighbor} in the multi-channel topology and has not to be covered. 
This mechanically reduces the overhead.
The probability of such configuration is smaller with dynamic interfaces since
we increase the probability that a pair of nodes has at least one channel in common at a given instant.

Finally, the Mixed/Pseudo-Random \& Adaptive strategy presents the worst
overhead, because it uses only one static interface, which reduces the possibilities to re-use one single transmission to cover several neighbors.

We have also evaluated fairness between different channels with the Jain Index (Figure~\ref{fig:nbnodes_jain}).
Mixed Interfaces with Common/Adaptive Channel assignment result in the Jain index of about 0.08.
Indeed, only the control channel (1 of the 12 channels) is used for broadcast leading to high unfairness.

Other strategies lead to almost perfect fairness: they efficiently spread the broadcast traffic through orthogonal channels reducing the risk of congestion.
In particular, the Static/Common strategy does not use a single control channel, leading to a good fairness.

\subsection{Density}

\begin{figure}[t]
\begin{center}
	\includegraphics[width=0.7\linewidth]{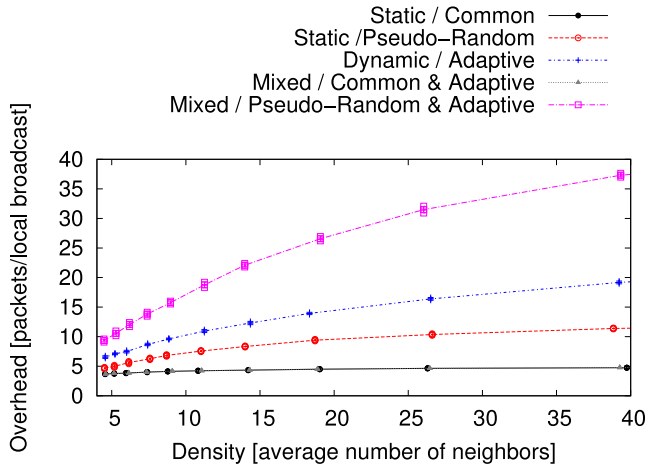}
	\caption{Impact of the density on the overhead, 200 nodes, 3 interfaces, 12 channels}
	\label{fig:density_overhead}
\end{center}
\end{figure}

We have also evaluated the impact of the density on the overhead while
maintaining the number of nodes constant (cf. Figure~\ref{fig:density_overhead}). 
Only Static/Common and Mixed/Common \& Adaptive strategies have the same overhead, which is perfectly
scalable, because they have a common channel set. 

The overhead created by Algorithm~\ref{algo:greedyStatic} applied to the
Static/Pseudo-Random slightly increases with the density: the greedy approach
succeeds to better take advantage of transmissions.
This growth is more important when we use dynamic interfaces as more timeslots are necessary to cover the interface schedule of new neighbors.

The Mixed/Pseudo-Random \& Adaptive strategy keeps on presenting the worst overhead since only one static interface is used for reception limiting the possibilities to use one single packet to cover several neighbors.

In conclusion, our greedy strategies are particularly efficient in minimizing the overhead when the density is large.

\subsection{Number of interfaces}
\begin{figure}
\begin{center}
	\includegraphics[width=0.7\linewidth]{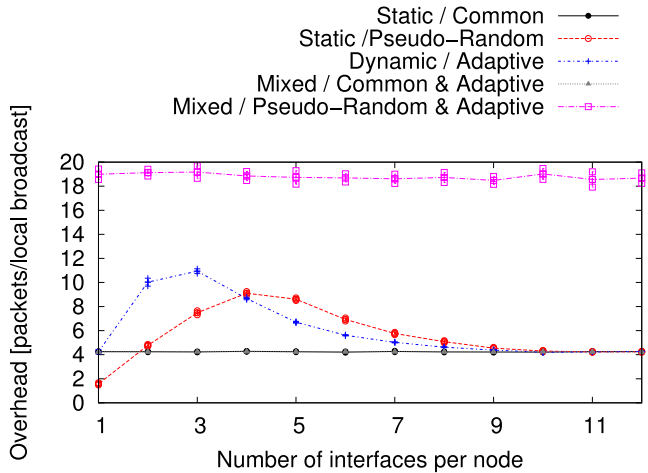}
	\caption{Impact of the number of interfaces on the overhead, 200 nodes, 10 avg. neighbors, 12 channels}
	\label{fig:nbint_overhead}
\end{center}
\end{figure}

We have also considered the influence of the number of interfaces (cf. Figure~\ref{fig:nbint_overhead}) on the overhead. 

The Dynamic/Adaptive and the Static/Pseudo-Random  strategies tend to have initially a growing overhead: the number of neighbors to be covered increases since they have more chance to have a common timeslot.
Then, the overhead decreases when it exceeds a threshold since the probability of having different neighbors that use the same channel increases with the number of interfaces.
The Dynamic/Adaptive begins to be more attractive when the number of interfaces is large compared to the number of channels ($> 3$ interfaces).
Finally, for a very large number of interfaces ($>8$), these strategies tend to be similar to the common channel strategies. 

The strategies using a common channel for broadcast are not impacted by the number of interfaces. 
Besides, the Mixed/Pseudo-Random \& Adaptive strategy presents also a constant overhead since the unique receiving interface keeps on being the bottleneck.

\subsection{Impact of threshold $p_{cover_{min}}$}
\begin{figure}
\begin{center}
	\includegraphics[width=0.7\linewidth]{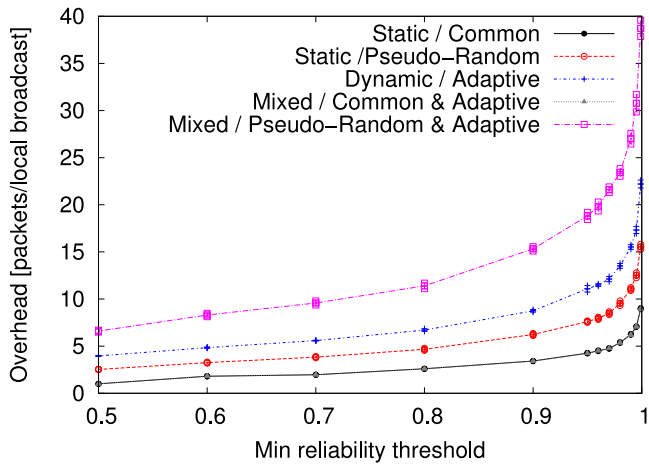}
	\caption{Impact of threshold $p_{cover_{min}}$  on the overhead, 200 nodes, 10 avg. neighbors, 3 interfaces, 12 channels}
	\label{fig:reliability_overhead}
\end{center}
\end{figure}

Finally, we have measured the impact of threshold $p_{cover_{min}}$ on the overhead in Figure~\ref{fig:reliability_overhead}.
When $p_{cover_{min}}=0.5$, each neighbor is covered when the transmitter sends one single copy: we discard radio links with a larger PER, considering them unreliable. 
Thus, strategies with a common channel set required only one broadcast transmissions to cover all the neighbors.

When $p_{cover_{min}}$ increases, the overhead becomes larger: neighbors with a
large packet error probability may require the transmission of several copies.
However, we can note that all the strategies follow the same trend.  
The overhead becomes prohibitive when we require very large $p_{cover_{min}}$ (e.g. $\approx 0.99$).
Thus, the network protocols have to cope with broadcast unreliability.
In particular, they should work in a self-stabilizing manner: even if some neighbors do not receive a particular broadcast packet, the protocol must work properly.


\section{Conclusion and Future Work}
\label{section:conclusion}

We have proposed algorithms for local broadcast in multi-channel
multi-interface wireless mesh networks.
In particular, they can cope with dynamic interfaces without a common control channel. 
To the best of our knowledge, these algorithms are the first ones to cope with deafness in this situation. 
Simulations show that all the strategies have an acceptable overhead and the load is fairly distributed among channels when the Common Channel Set strategy is not used.
A greedy approach is particularly efficient to take advantage of the broadcast
nature of transmissions.  

We plan to study how we can deal with multiple rates: different bit rates may
cover a different set of neighbors with different PER.  
We also plan to adapt the proposed strategies to dynamic conditions adopting an opportunistic approach.
Besides, we aim at optimizing the delay, e.g. consider the question of which
timeslot would present the best trade-off between the delay and the overhead
when we use dynamic interfaces.

\section*{Acknowledgments}

This work was supported in part by the French Government (MESR) and the Competitive Clusters Minalogic and System@tic under contracts FUI SensCity as well as the French National Research Agency (ANR) project ARESA2 ANR-09-VERS-017.

\bibliographystyle{unsrt}

\bibliography{biblio}

\end{document}